\documentclass[preprint,superscriptaddress,aps]{revtex4}
\usepackage{amsfonts}
\usepackage{graphics}
\usepackage{graphicx}
\usepackage{slashed}
\newcommand{\be}{\begin{equation}}
\newcommand{\ee}{\end{equation}}
\newcommand{\en}{\end{equation}}
\newcommand{\ba}{\begin{eqnarray}}
\newcommand{\ea}{\end{eqnarray}}
\newcommand{\bea}{\begin{eqnarray}}
\newcommand{\eea}{\end{eqnarray}}
\newcommand{\pa}{\partial}
\def\pls{\partial\!\!\!/}

\def\bs{b\!\!\!/}

\def\g{\gamma}

\def\eff{\mathrm{eff}}

\begin{document}

\title{Higher-derivative Lorentz-breaking terms in extended QED at the finite temperature}
\author{A. Celeste}
\affiliation{Instituto de F\'\i sica, Universidade Federal de Alagoas, 57072-270, Macei\'o, Alagoas, Brazil}
\email{aceleste,tmariz@fis.ufal.br}
\author{T. Mariz}
\affiliation{Instituto de F\'\i sica, Universidade Federal de Alagoas, 57072-270, Macei\'o, Alagoas, Brazil}
\email{aceleste,tmariz@fis.ufal.br}
\author{J. R. Nascimento}
\affiliation{Departamento de F\'{\i}sica, Universidade Federal da Para\'{\i}ba\\
 Caixa Postal 5008, 58051-970, Jo\~ao Pessoa, Para\'{\i}ba, Brazil}
\email{jroberto, petrov@fisica.ufpb.br}
\author{A. Yu. Petrov}
\affiliation{Departamento de F\'{\i}sica, Universidade Federal da Para\'{\i}ba\\
 Caixa Postal 5008, 58051-970, Jo\~ao Pessoa, Para\'{\i}ba, Brazil}
\email{jroberto, petrov@fisica.ufpb.br}
\begin{abstract}
In this paper we discuss finiteness and ambiguities of the higher-derivative Lorentz-breaking terms in extended QED with a magnetic coupling at the finite temperature. We find that, beside of the higher-derivative Carroll-Field-Jackiw-like term and Myers-Pospelov term, many extra terms arise in a finite temperature case but these terms vanish in high temperature limit. Moreover, the contributions for the nonminimal coupling will dominate at large temperatures.
\end{abstract}

\maketitle

\section{Introduction}

The higher-derivative terms, originally introduced to achieve the renormalizability of the quantum gravity \cite{Stelle}, now are studied within many contexts including Lorentz symmetry breaking. There are lot of reasons of interest to such terms -- first, they improve the renormalization properties of the theory, second, they naturally emerge within the derivative expansion of the effective action, third, they are probably related with the anomalies. It should be noted so that one of the first known Lorentz-breaking terms contains higher derivatives, it is the four-dimensional gravitational Chern-Simons term \cite{JaPi}. Therefore, the problem of studying the higher-derivative Lorentz-breaking terms in different field theory models seems to be very natural. The first example of such a term has been proposed in \cite{MP} within the phenomenological context, and in \cite{KostMew} the generic features of higher-derivative Lorentz-breaking terms within different extensions of QED have been discussed. Some issues related to these terms at the classical level, such as dispersion relations and unitarity and causality problems, have been also studied in \cite{MP,Reyes}. Moreover, while these terms in many papers have been generated in usual, zero temperature case \cite{HDLV0,HDLV}, it is natural to study their high-temperature dynamics. Such a study has been carried out for some extended versions of QED \cite{HDLV_T}. However, it is interesting to carry out this study for the specific extended QED with a magnetic coupling known to generated two different ambiguities \cite{magnQED}. So, our aim will consist in generating the higher-derivative terms at the finite temperature for this model. We show that, in the high temperature limit, the contribution generated by the essentially nonminimal sector will dominate.

The structure of the paper looks like follows. In the section 2 we introduce the classical action of our theory and write down the generic form of the one-loop effective action. In the section 3 we carry out the calculation of the three-derivative contribution to the two-point function. The Summary is devoted to the discussion of the results. In the Appendix, the temperature-dependent parameters of the one-loop corrections are listed.  

\section{Extended QED with magnetic coupling}

We start with the extended QED with magnetic coupling \cite{HDLV,MCN}: 
\be
\label{mcn}
{\cal L}=\bar{\psi}\left[i \pls- \gamma^{\mu}(eA_{\mu}+g\epsilon_{\mu\nu\lambda\rho}F^{\nu\lambda}b^{\rho}) - m -  \gamma_{5}\bs\right]\psi-\frac{1}{4}F_{\mu\nu}F^{\mu\nu}.
\ee
Here we suggest that the Lorentz symmetry breaking is introduced through a constant vector $b_{\mu}$. The $F_{\mu\nu}=\partial_{\mu}A_{\nu}-\partial_{\nu}A_{\mu}$ is the usual stress tensor constructed on the base of the gauge field $A_{\mu}$. Many impacts of the magnetic coupling have been studied in \cite{magnQED,aether,nmin}. 

As usual, we can express the one-loop effective action $S_\eff[b,A]$ of the gauge field $A_{\mu}$ in terms of the following functional trace:
\be
\label{det}
S_\eff[b,A]=-i\,{\rm Tr}\,\ln(\slashed{p}-\gamma^{\mu}\tilde{A}_{\mu}- m - \gamma_5 \bs),
\ee
where 
\bea
\label{tilde}
\tilde{A}_{\mu}=eA_{\mu}+g\epsilon_{\mu\nu\lambda\rho}F^{\nu\lambda}b^{\rho}.
\eea
We expand the trace (\ref{det}) in the power series:
\be
\label{ea}
S'_\eff[b,A]=i\,{\rm Tr} \sum_{n=1}^{\infty}\frac1n
\Biggl[\frac1{\slashed{p}- m - \gamma_5 \bs}\, \g^{\mu}\tilde{A}_{\mu}\Biggr]^n.
\ee 
We want to study the contributions of the second order in $\tilde{A}_{\mu}$ which can be read off from
\be
S_\eff^{(2)}[b,A]=\frac{i}{2}{\rm Tr}\frac{1}{ \slashed{p}- m - \gamma_5 \bs}\;\g^{\mu}\tilde{A}_{\mu}\;\frac{1}{ \slashed{p}- m - \gamma_5 \bs}\;\g^{\nu}\tilde{A}_{\nu},
\ee
or, as is the same,
\be
S_\eff^{(2)}[b,A]=\frac{i}{2}\int d^4x\, \Pi_b^{\mu\nu}\tilde{A}_{\mu}\tilde{A}_{\nu},
\ee
where
\be\label{Pib}
\Pi_b^{\mu\nu}={\rm tr}\int\frac{d^4l}{(2\pi)^4}\frac{1}{\slashed{l}-m-\gamma_5\slashed{b}}\gamma^\mu\frac{1}{\slashed{l}-i\slashed{\partial}-m-\gamma_5\slashed{b}}\gamma^\nu.
\ee
Our aim here consists in expansion of this expression up to the third order in derivatives.
Repeating the arguments from \cite{HDLV}, we can show that the corresponding term must be of third order in the Lorentz-breaking vector $b_{\mu}$, with there will be three different contributions, where the number of insertions of the vector $b^{\mu}$ into the propagators is equal to one, two or three, which is equivalent to two, one or zero ``magnetic'' vertices $g\bar{\psi}\epsilon_{\mu\nu\lambda\rho}\gamma^{\mu}F^{\nu\lambda}b^{\rho}\psi$, respectively. 

\section{Higher-derivative quantum corrections}

So, now let us find explicitly the third-derivative quantum corrections.
First, we consider a contribution characterized by two nonminimal vertices. Using the above equation (\ref{Pib}), we must calculate the contribution given by 
\ba\label{I1}
\Pi_{b1}^{\mu\nu}(p) &=& {\rm tr}\int\frac{d^4l}{(2\pi)^4}S(l)\gamma_5\slashed{b}S(l)\gamma^\mu S(l-p)\gamma^\nu \nonumber\\
&&+{\rm tr}\int\frac{d^4l}{(2\pi)^4}S(l)\gamma^\mu S(l-p)\gamma_5\slashed{b}S(l-p)\gamma^\nu,
\ea
with $S(l)=(\slashed{l}-m)^{-1}$, where we taken into account that $i\partial\to p$, after a Fourier transform. This contribution corresponds to Fig. 1.

\begin{figure}[ht] 
\includegraphics[scale=0.6]{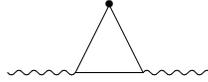}
\caption{\small{Contributions with two nonmimimal vertices.}}
\label{figure1}
\end{figure}
 
This expression is superficially divergent. It has been calculated in many papers (some related issues have been discussed in \cite{list}) in the context of the perturbative generation of the Carroll-Field-Jackiw (CFJ) term, since to obtain our higher-derivative contribution from this sector one should simply repeat the calculations of the CFJ term and further replace $A_{\mu}\to\epsilon_{\mu\nu\lambda\rho}b^{\nu}F^{\lambda\rho}$ in the external legs, with the trace of the product of propagators and Dirac matrices continues to be the same, given by (\ref{I1}).
The result for this contribution to the self-energy tensor looks like $\Pi_{b1}^{\mu\nu}(p) \propto \epsilon^{\mu\nu\lambda\rho}p_\lambda k_\rho$, where the $p_{\mu}$ is an external momentum, and the $k_{\rho}$ is a constant vector which turns out to be finite. 
Within its calculation we follow the regularization prescription introduced in \cite{Vel}  and applied to the calculation of the CFJ term in \cite{many}.
Its explicit form can be read off from the expression  
\begin{eqnarray}
\Pi_{b1}^{\mu\nu}(p) &=& 4i\int\,\frac{d^4l}{(2\pi)^4}\,\frac{1}{(l^2-m^2)^3}[3(l^2-m^2)\epsilon^{\mu\nu\lambda\rho}p_\lambda b_\rho - 4l^\mu l_\alpha \epsilon^{\alpha\nu\lambda\rho}p_\lambda b_\rho \nonumber\\
&& - 4l^\nu l_\alpha \epsilon^{\mu\alpha\lambda\rho}p_\lambda b_\rho - 4 (l\cdot p) l_\alpha \epsilon^{\mu\nu\alpha\rho} b_\rho].
\end{eqnarray}
To find this vector in the finite temperature case, one should follow the Matsubara formalism. To do it, one carries the Wick rotation and discretization of the zero component by the rule $l_0\to 2\pi T(n+\frac{1}{2})$, for the integer $n$ and replaces integration over $dl_0$ by summation over $n$. As a result, one finds $\Pi_{b1}^{\mu\nu}(p)$ as a function of $\xi$, with $\xi=\frac{m}{2\pi T}$:
\begin{eqnarray}\label{pi15}
\Pi_{b1}^{\mu\nu}(p) &=& A_1(\xi) \left(\delta^{\mu 0}\epsilon^{0\nu\lambda\rho}p_E^\lambda b_E^\rho+\delta^{\nu 0}\epsilon^{\mu0\lambda\rho}p_E^\lambda b_E^\rho+\epsilon^{\mu\nu0\rho}p_0 b_E^\rho\right) = A_1(\xi) \epsilon^{\mu\nu\lambda i} p_E^\lambda b_E^i,
\end{eqnarray}
where
\begin{equation}
A_1(\xi)=\int_{|\xi|}^{\infty} dz(z^{2}-\xi^{2})^{1/2}\mathrm{sech}^{2}{(\pi z)}\tanh{(\pi z)}.
\end{equation}
Note that
\begin{equation}
A_1(\xi \rightarrow 0) \to \int_{0}^{\infty} dz\ z\ \mathrm{sech}^{2}{(\pi z)}\tanh{(\pi z)} = \frac{1}{2\pi^{2}},
\end{equation}
so, it does not vanish. At the same time, in the zero temperature limit $\xi\to\infty$, one gets $A_1(\xi)|_{\xi\to\infty}\to 0$, so, at the zero temperature this contribution vanishes which matches the result found in \cite{many}.
The corresponding contribution to the effective action is
\be\label{s1}
S_{AA,1}=2g^2A_1(\xi)\int d^4x\,  \left[b^\alpha F_{\alpha\mu}(b\cdot\partial)b_i\epsilon^{\mu\nu\lambda i}F_{\nu\lambda}+b^2b_i\epsilon^{\mu\nu\lambda i}A_\mu\Box F_{\nu\lambda}\right].
\ee

Now, let us consider another two contributions of third order in $b^{\mu}$, which essentially require consideration of the minimal coupling. First, one has the contribution to the effective Lagrangian involving one vertex with minimal coupling, and another vertex with nonminimal coupling, and two insertions in the propagators, given by Feynman diagrams depicted at Fig.2. Therefore, the contribution we need to calculate from Eq.~(\ref{Pib}), is given by
\ba
\label{pib2}
\Pi_{b2}^{\mu\nu}(p) &=& {\rm tr}\int\frac{d^4l}{(2\pi)^4}S(l)\gamma_5\slashed{b}S(l)\gamma_5\slashed{b}S(l)\gamma^\mu S(l-p)\gamma^\nu \nonumber\\
&&+{\rm tr}\int\frac{d^4l}{(2\pi)^4}S(l)\gamma_5\slashed{b}S(l)\gamma^\mu S(l-p)\gamma_5\slashed{b}S(l-p)\gamma^\nu \nonumber\\
&&+{\rm tr}\int\frac{d^4l}{(2\pi)^4}S(l)\gamma^\mu S(l-p)\gamma_5\slashed{b}S(l-p)\gamma_5\slashed{b}S(l-p)\gamma^\nu.
\ea

\begin{figure}[ht] 
\includegraphics[scale=0.6]{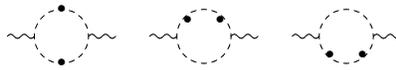}
\caption{\small{Contributions with one nonminimal vertex.}}
\label{figure2}
\end{figure}

We expand it up to the second order in the external momentum $p_{\mu}$. As a result, we find that
the expression of Eq.(\ref{pib2}) can then be written as
\begin{equation}
S_{AA,2}(p)= ie^2A_{\mu}(-p)\epsilon_{\nu\lambda\rho\sigma}\Pi_{b2}^{\mu\nu}|(p)b^{\lambda}F^{\rho\sigma}(p) + {\cal O}(p^3),
\end{equation}
where $\Pi_{b2}|$ is the contribution to the $\Pi_{b2}$ given by (\ref{pib2}) involving only the second order in external momentum $p$. It has been calculated in details in \cite{ouraether}, so, we merely quote the result:
\begin{eqnarray}\label{sumprojs}
\Pi^{\mu\nu}_{b2}&=&A_2(\xi)((b_E\cdot p_E)^2 \delta^{\mu\nu}+p_E^2 b_E^\mu b_E^\nu-(b_E\cdot p_E)b_E^\mu p_E^\nu-(b_E\cdot p_E)p_E^\mu b_E^\nu-b_E^2p_E^2 \delta^{\mu\nu}+b_E^2 p_E^\mu p_E^\nu) \nonumber\\
&&+B_2(\xi)(b_0^2p_0^2\delta^{\mu\nu}+b_0^2p_E^2\delta^{\mu0}\delta^{\nu0}-b_0^2p_0 p_E^\mu\delta^{\nu0}-b_0^2p_0 \delta^{\mu0}p_E^\nu) \nonumber\\
&&+C_2(\xi)(b_E^2p_0^2\delta^{\mu\nu}+b_E^2p_E^2\delta^{\mu0}\delta^{\nu0}-b_E^2p_0\delta^{\mu0}p_E^\nu-b_E^2p_0p_E^\mu\delta^{\nu0}) \nonumber\\
&&+D_2(\xi)(b_0^2p_E^2\delta^{\mu\nu}-b_0^2p_E^\mu p_E^\nu),
\end{eqnarray}
where
\begin{eqnarray}
A_2(\xi)&=&-\frac{1}{6m^2\pi^{2}}-\int_{|\xi|}^{\infty}dz\frac{(\xi^{2}-2z^{2})}{6m^2(z^2-\xi^2)^{1/2}}{\rm sech}^2(\pi z)\tanh(\pi z),
\end{eqnarray}
\begin{eqnarray}
B_2(\xi)&=& -C_2(\xi) = \int_{|\xi|}^{\infty}dz\frac{\pi^{2}\xi^{2}(z^2-\xi^2)^{1/2}}{12m^2}{\rm sech}^5(\pi z)(\sinh(3\pi z)-11\sinh(\pi z)),
\end{eqnarray}
and
\begin{eqnarray}
D_2(\xi)&=& \int_{|\xi|}^{\infty}dz \frac{\xi^{2}}{6m^2(z^2-\xi^2)^{1/2}} {\rm sech}^{2}(\pi z)\tanh(\pi z).
\end{eqnarray}
We observe that in the high temperature limit, $\xi\to0$, all the above coefficients vanish.

So, we can write the following result for the higher-derivative contribution from this sector:
\bea
\label{s2}
S_{AA,2}&=&-A_2(\xi)\epsilon_{\nu\rho\lambda\sigma}A^{\nu}[(b\cdot\partial)^2-b^2\Box]b^{\rho}F^{\lambda\sigma}-D_2(\xi)b^2_0\epsilon_{\nu\rho\lambda\sigma}A^{\nu}b^{\rho}\Box F^{\lambda\sigma}\\
&&-B_2(\xi)(b^2_0-b^2_E)A^{\mu}(\epsilon_{\mu\rho\lambda\sigma}\pa^2_0+\pa_{\mu}\pa_0\epsilon_{0\rho\lambda\sigma})b^{\rho}F^{\lambda\sigma}-B_2(\xi)(b^2_0-b^2_E)
A^0\epsilon_{0\rho\lambda\sigma}b^{\rho}\Box F^{\lambda\sigma}.\nonumber
\eea
This expression can be verified to be gauge invariant at the finite temperature. We see that this expression involves the higher-derivative CFJ-like term (proportional to $D_2(\xi)$), the Myers-Pospelov term (proportional to $A_2(\xi)$) and some extra contributions which can be treated as generalizations of these terms (proportional to $B_2(\xi)$). 

It remains to consider only the contribution to the effective Lagrangian involving both vertices with minimal coupling which requires three insertions of the $\gamma_5\bs$ into the propagator. The corresponding Feynman diagrams are depicted at Fig. 3.

\begin{figure}[ht] 
\includegraphics[scale=0.6]{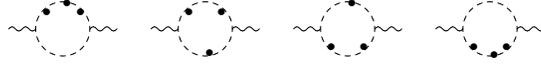}
\caption{\small{Contributions with minimal vertices only.}}
\label{figure2}
\end{figure}

Thus, the contribution to the self-energy tensor from this sector looks like
\ba
\Pi_{b3}^{\mu\nu}(p) &=& {\rm tr}\int\frac{d^4l}{(2\pi)^4}S(l)\gamma_5\slashed{b}S(l)\gamma_5\slashed{b}S(l)\gamma_5\slashed{b}S(l)\gamma^\mu S(l-p)\gamma^\nu \nonumber\\
&&+{\rm tr}\int\frac{d^4l}{(2\pi)^4}S(l)\gamma_5\slashed{b}S(l)\gamma_5\slashed{b}S(l)\gamma^\mu S(l-p)\gamma_5\slashed{b}S(l-p)\gamma^\nu \nonumber\\
&&+{\rm tr}\int\frac{d^4l}{(2\pi)^4}S(l)\gamma_5\slashed{b}S(l)\gamma^\mu S(l-p)\gamma_5\slashed{b}S(l-p)\gamma_5\slashed{b}S(l-p)\gamma^\nu \nonumber\\
&&+{\rm tr}\int\frac{d^4l}{(2\pi)^4}S(l)\gamma^\mu S(l-p)\gamma_5\slashed{b}S(l-p)\gamma_5\slashed{b}S(l-p)\gamma_5\slashed{b}S(l-p)\gamma^\nu.
\ea

The result at the finite temperature is given by
\ba
\Pi_{b3}^{\mu\nu}(p) &=& A_3(\xi) \epsilon^{\mu\nu\lambda\rho} b_E^\lambda p_E^\rho b_E^2 p_E^2 + B_3(\xi) \epsilon^{\mu\nu\lambda\rho} b_E^\lambda p_E^\rho (b_E\cdot p_E)^2 + C_3(\xi) \epsilon^{\mu\nu\lambda\rho} b_E^\lambda p_E^\rho b_E^2 p_0^2 \nonumber \\
&& + D_3(\xi) \epsilon^{\mu\nu\lambda\rho} b_E^\lambda p_E^\rho b_0^2 p_E^2 +  E_3(\xi) \epsilon^{\mu\nu\lambda\rho} b_E^\lambda p_E^\rho (b_E\cdot p_E) b_0 p_0  +  F_3(\xi) \epsilon^{\mu\nu\lambda\rho} b_E^\lambda p_E^\rho b_0^2 p_0^2 \nonumber \\
&& + G_3(\xi) \epsilon^{\mu\nu\lambda0} b_E^\lambda p_0 b_E^2 p_E^2 + H_3(\xi) \epsilon^{\mu\nu\lambda0} b_E^\lambda p_0 (b_E\cdot p_E)^2 + I_3(\xi) \epsilon^{\mu\nu\lambda0} b_E^\lambda p_0 b_E^2 p_0^2 \nonumber \\
&& + J_3(\xi) \epsilon^{\mu\nu\lambda0} b_E^\lambda p_0 b_0^2 p_E^2 +  K_3(\xi) \epsilon^{\mu\nu\lambda0} b_E^\lambda p_0 (b_E\cdot p_E) b_0 p_0  +  L_3(\xi) \epsilon^{\mu\nu\lambda0} b_E^\lambda p_0 b_0^2 p_0^2 \nonumber \\
&& + M_3(\xi) \epsilon^{\mu\nu\lambda0} b_E^\lambda b_0 (b_E\cdot p_E) p_E^2 + N_3(\xi) \epsilon^{\mu\nu0\rho} b_0 p_E^\rho b_E^2 p_E^2 + O_3(\xi) \epsilon^{\mu\nu0\rho} b_0 p_E^\rho b_E^2 p_0^2 \nonumber \\
&& + P_3(\xi) \epsilon^{\mu\nu0\rho} p_0 p_E^\rho (b_E\cdot p_E) b_E^2 + Q_3(\xi) (\epsilon^{\mu0\lambda\rho} \delta^{\nu0}  + \epsilon^{0\nu\lambda\rho} \delta^{\mu0})b_E^\lambda p_E^\rho b_E^2 p_E^2 \nonumber \\
&& + R_3(\xi) (\epsilon^{\mu0\lambda\rho} \delta^{\nu0}(b_E\cdot p_E) + \epsilon^{0\nu\lambda\rho} \delta^{\mu0}(b_E\cdot p_E) - \epsilon^{\mu0\lambda\rho} b_0 p_E^\nu - \epsilon^{0\nu\lambda\rho} b_0 p_E^\mu)b_E^\lambda p_E^\rho(b_E\cdot p_E) \nonumber \\
&& + S_3(\xi) (\epsilon^{\mu0\lambda\rho} \delta^{\nu0}(b_E\cdot p_E) + \epsilon^{0\nu\lambda\rho} \delta^{\mu0}(b_E\cdot p_E) - \epsilon^{\mu0\lambda\rho} b_0 p_E^\nu - \epsilon^{0\nu\lambda\rho} b_0 p_E^\mu)b_E^\lambda p_E^\rho b_0 p_0 \nonumber \\
&& + T_3(\xi) (\epsilon^{\mu0\lambda\rho} \delta^{\nu0} + \epsilon^{0\nu\lambda\rho} \delta^{\mu0})b_E^\lambda p_E^\rho b_0^2 p_0^2 + U_3(\xi) (\epsilon^{\mu0\lambda\rho} \delta^{\nu0} + \epsilon^{0\nu\lambda\rho} \delta^{\mu0})b_E^\lambda p_E^\rho b_0^2 p_E^2 \nonumber \\
&& + V_3(\xi) (\epsilon^{\mu0\lambda\rho} \delta^{\nu0} + \epsilon^{0\nu\lambda\rho} \delta^{\mu0})b_E^\lambda p_E^\rho b_E^2 p_0^2 + W_3(\xi) (\epsilon^{\mu0\lambda\rho} p_E^\nu + \epsilon^{0\nu\lambda\rho} p_E^\mu)b_E^\lambda p_E^\rho b_E^2 p_0 \nonumber \\
&& + X_3(\xi) (\epsilon^{\mu0\lambda\rho} b_E^\nu + \epsilon^{0\nu\lambda\rho} b_E^\mu)b_E^\lambda p_E^\rho b_0 p_E^2 + Y_3(\xi) (\epsilon^{\mu0\lambda\rho} b_E^\nu + \epsilon^{0\nu\lambda\rho} b_E^\mu)b_E^\lambda p_E^\rho (b_E\cdot p_E) p_0 \nonumber \\
&& + Z_3(\xi) (\epsilon^{\mu0\lambda\rho} b_E^\nu + \epsilon^{0\nu\lambda\rho} b_E^\mu)b_E^\lambda p_E^\rho b_0 p_0^2,
\ea
where the temperature-dependent constants $A_3\ldots Z_3$ are given in Appendix. We note that in the high temperature limit, $\xi\to0$, all these coefficients vanish. Also, it is easy to verify that this contribution is transversal and hence gauge independent (indeed, its contraction with $p_{\mu}$ or $p_{\nu}$ gives 0).

In principle, this term can be transformed to the coordinate space. It yields the following contribution to the effective action:
\ba
\label{s3}
S_{AA,3} &=& -\frac{1}{2}A_{\mu}\Big(\epsilon^{\mu\nu\lambda\rho}\Big[A_3(\xi)   b^2 \Box + B_3(\xi)  (b\cdot\partial)^2 + C_3(\xi)  b^2 \partial_0^2 \nonumber \\
&& + D_3(\xi) b_0^2 \Box +  E_3(\xi)  (b\cdot \partial) b_0 \partial_0  +  F_3(\xi) b_0^2 \partial_0^2\Big]b_\lambda \partial_\rho \nonumber \\
&& +\epsilon^{\mu\nu\lambda0}b_{\lambda}\Big[ G_3(\xi) \partial_0 b^2 \Box + H_3(\xi) \partial_0 (b\cdot \partial)^2 + I_3(\xi)  \partial_0 b^2 \partial_0^2  + 
J_3(\xi)\partial_0 b_0^2 \Box \nonumber \\&& +  K_3(\xi)   \partial_0 (b\cdot \partial) b_0 \partial_0  +  L_3(\xi)  \partial_0 b_0^2 \partial_0^2  + M_3(\xi) b_0 (b\cdot \partial) \Box\Big] \nonumber \\&&
+ \epsilon^{\mu\nu0\rho}b^2\partial_{\rho}\Big[N_3(\xi) b_0 \Box + O_3(\xi) b_0 \partial_0^2 
+  P_3(\xi) \partial_0 (b\cdot \partial) \Big]  \nonumber\\ && +
\Big[Q_3(\xi) (\epsilon^{\mu0\lambda\rho} \delta^{\nu0}  + \epsilon^{0\nu\lambda\rho} \delta^{\mu0})b^2  \Box  +  \Big(R_3(\xi)(b\cdot \partial) + S_3(\xi)  b_0 \partial_0\Big)\nonumber\\&& \times
(\epsilon^{\mu0\lambda\rho} \delta^{\nu0}(b\cdot \partial) + \epsilon^{0\nu\lambda\rho} \delta^{\mu0}(b\cdot \partial) - \epsilon^{\mu0\lambda\rho} b_0 \partial^\nu - \epsilon^{0\nu\lambda\rho} b_0 \partial^\mu) \nonumber \\
&& + T_3(\xi)(\epsilon^{\mu0\lambda\rho} \delta^{\nu0} + \epsilon^{0\nu\lambda\rho} \delta^{\mu0}) b_0^2 \partial_0^2 + U_3(\xi) (\epsilon^{\mu0\lambda\rho} \delta^{\nu0} + \epsilon^{0\nu\lambda\rho} \delta^{\mu0}) b_0^2 \Box \nonumber \\
&& + V_3(\xi) (\epsilon^{\mu0\lambda\rho} \delta^{\nu0} + \epsilon^{0\nu\lambda\rho} \delta^{\mu0}) b^2 \partial_0^2 + W_3(\xi) (\epsilon^{\mu0\lambda\rho} \partial^\nu + \epsilon^{0\nu\lambda\rho} \partial^\mu) b^2 \partial_0 \nonumber \\
&& + X_3(\xi) (\epsilon^{\mu0\lambda\rho} b^\nu + \epsilon^{0\nu\lambda\rho} b^\mu) b_0 \Box + Y_3(\xi) (\epsilon^{\mu0\lambda\rho} b^\nu + \epsilon^{0\nu\lambda\rho} b^\mu) (b\cdot \partial) \partial_0 \nonumber \\
&& + Z_3(\xi) (\epsilon^{\mu0\lambda\rho} b^\nu + \epsilon^{0\nu\lambda\rho} b^\mu) b_0 \partial_0^2\Big]b_{\lambda}\partial_{\rho}\Big)A_{\nu}.
\ea
The final result is a sum of (\ref{s1},\ref{s2},\ref{s3}).
We emphasize again that it is finite and gauge invariant. We see that, besides of straightforward finite temperature generations of higher-derivative CFJ and Myers-Pospelov terms (in (\ref{s3}) these terms contribute to $A_3$, $B_3$ and $D_3$), we have a lot of new terms which has no analogues at zero temperature. The similar situation occurs in a previous paper \cite{ouraether} by some of us, where perturbative generation of the aether term has been performed at the finite temperature. 

\section{Summary}

Now, let us discuss our results. We have considered the perturbative generation of the three-derivative gauge-invariant term in the extended QED involving both minimal and nonminimal couplings in the finite temperature case. This term turns out to be gauge invariant and UV finite, reproducing a linear combination of the Myers-Pospelov term, known for the highly nontrivial manner of the propagating of solutions admitting for the rotation of plane of polarization of light \cite{MP}, the higher-derivative CFJ term, and some extra terms which have no zero-temperature analogues. An interesting observation consists in the fact that the terms generated by two and three insertions of the extra Lorentz-breaking vertex $\bs\gamma_5$ into the propagator vanish in the high temperature limit which apparently means that the nonminimal contribution becomes dominant in this limit.

{\bf Acknowledgements.}  This work was partially supported by Conselho
Nacional de Desenvolvimento Cient\'{\i}fico e Tecnol\'{o}gico (CNPq). The work by A. Yu. P. has been supported by the CNPq project No. 303438/2012-6.

\vspace*{3mm}

\centerline{\bf APPENDIX}

In this Appendix, we present the explicit results for the temperature-dependent factors $A_3\ldots Z_3$. 
\ba
A_3(\xi) &=& \frac{1}{9\pi^2m^4} + \int_{|\xi|}^\infty dz \frac{ \tanh(\pi  z)  \mathrm{sech}^4(\pi  z)}{72 m^4 (z^2-\xi^2)^{1/2}}\left(-25 \pi ^2 \xi ^4+4 \xi ^2+\left(25 \pi ^2 \xi ^2-8\right) z^2 \right. \nonumber \\
&&\left.+\left(5 \pi ^2 \xi ^4+4 \xi ^2-\left(5 \pi ^2 \xi ^2+8\right) z^2\right)  \cosh(2 \pi  z)\right),
\ea
\ba
B_3(\xi) &=& -\frac{4}{45\pi^2m^4} + \int_{|\xi|}^\infty dz \frac{\tanh(\pi  z) \mathrm{sech}^4(\pi  z)}{90 m^4 (z^2-\xi^2)^{1/2}}\left(15 \pi ^2 \xi ^4-\xi ^2 \left(15 \pi ^2 z^2+4\right) \right. \nonumber \\
&&\left.+\left(-3 \pi ^2 \xi ^4+\xi ^2 \left(3 \pi ^2 z^2-4\right)+8 z^2\right)  \cosh(2 \pi  z) +8 z^2\right),
\ea
\ba
C_3(\xi) &=& \int_{|\xi|}^\infty dz \frac{\pi ^2 \xi ^2 \left(7 \xi ^2-33 z^2\right)}{360 m^4 (z^2-\xi^2)^{1/2}} (\cosh(2 \pi  z) -5) \tanh(\pi  z) \mathrm{sech}^4(\pi  z),
\ea
\ba
D_3(\xi) &=& \int_{|\xi|}^\infty dz \frac{\pi ^2 \xi ^2 (z^2-\xi^2)^{1/2}}{36 m^4} (\cosh(2 \pi  z) -5) \tanh(\pi  z) \mathrm{sech}^4(\pi  z),
\ea
\ba
E_3(\xi) &=& \int_{|\xi|}^\infty dz \frac{\pi ^2 \xi ^2 \left(23 z^2-17 \xi ^2\right)}{180 m^4 (z^2-\xi^2)^{1/2}}(\cosh(2 \pi  z) -5) \tanh(\pi  z) \mathrm{sech}^4(\pi  z),
\ea
\ba
F_3(\xi) &=& -\int_{|\xi|}^\infty dz \frac{\pi ^2 \xi ^4}{90 m^4 (z^2-\xi^2)^{1/2}}(\cosh(2 \pi  z) -5) \tanh(\pi  z) \mathrm{sech}^4(\pi  z),
\ea
\ba
G_3(\xi) &=& -\int_{|\xi|}^\infty dz \frac{\pi ^2 \xi ^2 \left(2 \xi ^2-3 z^2\right)}{120 m^4 (z^2-\xi^2)^{1/2}}(\sinh(3 \pi  z) -11\sinh(\pi  z)) \mathrm{sech}^5(\pi  z),
\ea
\ba
H_3(\xi) &=& \int_{|\xi|}^\infty dz \frac{\pi ^2 \xi ^2 \left(\xi ^2-5 z^2\right)}{120 m^4 (z^2-\xi^2)^{1/2}}(\sinh(3 \pi  z) -11\sinh(\pi  z)) \mathrm{sech}^5(\pi  z),
\ea
\ba
I_3(\xi) &=& \int_{|\xi|}^\infty dz \frac{\pi ^2 \xi ^4 \tanh(\pi  z) \mathrm{sech}^6(\pi  z)}{144 m^4 (z^2-\xi^2)^{1/2}}\left(984 \pi ^2 \left(\xi ^2-z^2\right)+\left(8 \pi ^2 \left(\xi ^2-z^2\right)+1\right) \cosh(4 \pi  z) \right.\nonumber \\
&&\left.+8 \left(56 \pi ^2 (z-\xi ) (\xi +z)-1\right) \cosh(2 \pi  z)-9\right),
\ea
\ba
J_3(\xi) &=& -\int_{|\xi|}^\infty dz \frac{\pi ^2 \xi ^4}{120 m^4 (z^2-\xi^2)^{1/2}}(\sinh(3 \pi  z) -11\sinh(\pi  z)) \mathrm{sech}^5(\pi  z),
\ea
\ba
K_3(\xi) &=& \int_{|\xi|}^\infty dz \frac{\pi ^2 \xi ^4}{20 m^4 (z^2-\xi^2)^{1/2}}(\sinh(3 \pi  z) -11\sinh(\pi  z)) \mathrm{sech}^5(\pi  z),
\ea
\ba
L_3(\xi) &=& \int_{|\xi|}^\infty dz \frac{\pi ^4 \xi ^4 (z^2-\xi^2)^{1/2}}{36 m^4}(302\sinh(\pi  z) + \sinh(5 \pi  z) -57\sinh(3 \pi  z)) \nonumber \\
&&\times\mathrm{sech}^7(\pi  z),
\ea
\ba
M_3(\xi) &=& \int_{|\xi|}^\infty dz \frac{\pi ^2 \xi ^2 (z^2-\xi^2)^{1/2}}{120 m^4}(\sinh(3 \pi  z) -11\sinh(\pi  z)) \mathrm{sech}^5(\pi  z),
\ea
\ba
N_3(\xi) &=& -\int_{|\xi|}^\infty dz \frac{\pi ^2 \xi ^2 (z^2-\xi^2)^{1/2}}{48 m^4}(\sinh(3 \pi  z) -11\sinh(\pi  z)) \mathrm{sech}^5(\pi  z),
\ea
\ba
O_3(\xi) &=& -\int_{|\xi|}^\infty dz \frac{\pi ^2 \xi ^4}{24 m^4 (z^2-\xi^2)^{1/2}}(\sinh(3 \pi  z) -11\sinh(\pi  z)) \mathrm{sech}^5(\pi  z),
\ea
\ba
P_3(\xi) &=& \int_{|\xi|}^\infty dz \frac{\pi ^2 \xi ^2 \left(\xi ^2+3 z^2\right)}{240 m^4 (z^2-\xi^2)^{1/2}}(\sinh(3 \pi  z) -11\sinh(\pi  z)) \mathrm{sech}^5(\pi  z),
\ea
\ba
Q_3(\xi) &=& \int_{|\xi|}^\infty dz \frac{\pi ^2 \xi ^2 \left(\xi ^2-3 z^2\right)}{144 m^4 (z^2-\xi^2)^{1/2}}(\sinh(3 \pi  z) -11\sinh(\pi  z)) \mathrm{sech}^5(\pi  z),
\ea
\ba
R_3(\xi) &=& \int_{|\xi|}^\infty dz \frac{\pi ^2 \xi ^2 (z^2-\xi^2)^{1/2}}{180 m^4}(\sinh(3 \pi  z) -11\sinh(\pi  z)) \mathrm{sech}^5(\pi  z),
\ea
\ba
S_3(\xi) &=& \int_{|\xi|}^\infty dz \frac{\pi ^2 \xi ^4}{45 m^4 (z^2-\xi^2)^{1/2}}(\sinh(3 \pi  z) -11\sinh(\pi  z)) \mathrm{sech}^5(\pi  z),
\ea
\ba
T_3(\xi) &=& \int_{|\xi|}^\infty dz \frac{\pi ^4 \xi ^4 (z^2-\xi^2)^{1/2}}{36 m^4}(302\sinh(\pi  z) +\sinh(5 \pi  z) -57\sinh(3 \pi  z)) \nonumber \\
&&\times\mathrm{sech}^7(\pi  z),
\ea
\ba
U_3(\xi) &=& \int_{|\xi|}^\infty dz \frac{\pi ^2 \xi ^4}{72 m^4 (z^2-\xi^2)^{1/2}}(\sinh(3 \pi  z) -11\sinh(\pi  z)) \mathrm{sech}^5(\pi  z),
\ea
\ba
V_3(\xi) &=& \int_{|\xi|}^\infty dz \frac{\pi ^2 \xi ^4 \tanh(\pi  z) \mathrm{sech}^6(\pi  z)}{144 m^4 (z^2-\xi^2)^{1/2}}\left(8 \left(56 \pi ^2 \left(\xi ^2-z^2\right)+1\right) \cosh(2 \pi  z) \right. \nonumber\\
&&\left.+984 \pi ^2 (\xi +z) (z-\xi )+\left(8 \pi ^2 (z-\xi ) (\xi +z)-1\right) \cosh(4 \pi  z) +9\right),
\ea
\ba
W_3(\xi) &=& \int_{|\xi|}^\infty dz \frac{\pi ^2 \xi ^2 \left(33 z^2-17 \xi ^2\right)}{720 m^4 (z^2-\xi^2)^{1/2}}(\sinh(3 \pi  z) -11\sinh(\pi  z)) \mathrm{sech}^5(\pi  z),
\ea
\ba
X_3(\xi) &=& \int_{|\xi|}^\infty dz \frac{\pi ^2 \xi ^2 (z^2-\xi^2)^{1/2}}{72 m^4}(\sinh(3 \pi  z) -11\sinh(\pi  z)) \mathrm{sech}^5(\pi  z),
\ea
\ba
Y_3(\xi) &=& \int_{|\xi|}^\infty dz \frac{\pi ^2 \xi ^2  \left(5 \xi ^2-17 z^2\right)}{360 m^4 (z^2-\xi^2)^{1/2}}(\sinh(3 \pi  z) -11\sinh(\pi  z)) \mathrm{sech}^5(\pi  z),
\ea
and
\ba
Z_3(\xi) &=& \int_{|\xi|}^\infty dz \frac{\pi ^2 \xi ^4}{36 m^4 (z^2-\xi^2)^{1/2}}(\sinh(3 \pi  z) -11\sinh(\pi  z)) \mathrm{sech}^5(\pi  z).
\ea
As we already mentioned, all these constants vanish in the high temperature limit.

\end{document}